\documentclass[dvipdfmx]{PoS}
\usepackage{url}

\title{Hyperon single-particle potentials from QCD\\ on lattice}

\ShortTitle{Hyperon single-particle potentials form QCD on lattice}

\author{\speaker{Takashi Inoue}\\
        Nihon University, College of Bioresource Sciences, Fujisawa 252-0880, Japan\\
        Theoretical Research Division, Nishina Center, RIKEN, Wako 351-0198, Japan\\
        E-mail: \email{inoue.takashi@nihon-u.ac.jp}}

\author{for HAL QCD Collaboration}

\abstract{
We study single-particle potential of hyperons in nuclear medium starting from QCD.
First we carry out lattice QCD numerical simulation to extract baryon-baryon interactions from QCD by means of the HAL QCD method. 
We employ a full QCD gauge configuration ensemble at almost physical point
so that hadron masses are nearly physical, {\it e.g.} pion mass is 146 MeV, kaon mass is 525 MeV, and nucleon mass is 956 MeV.
Then, with some simplifications, we apply the obtained hyperon interactions to the Brueckner-Hartree-Fock theory
and calculate single-particle potential of hyperons in nuclear medium $U_{Y}(\rho,k)$.
For the symmetric nuclear matter at the normal nuclear matter density,
we obtain $U_{\Lambda}(\rho_0,0)=-33$~MeV, $U_{\Sigma}(\rho_0,0)=+11$~MeV, and  $U_{\Xi}(\rho_0,0)=-6$~MeV.
These results are qualitatively compatible with values suggested from experiments.
This success is remarkable and encouraging because we are trying to reveal nature of baryon-baryon interactions starting from QCD,
and this agreement proves that our approach is essentially correct. 
}

\FullConference{The 26th International Nuclear Physics Conference\\
		11-16 September, 2016\\
		Adelaide, Australia}

\begin{document}

\def \etal{{\it et al.\,}}
\def \etc{{\it etc.\,}}
\def \ie{{\it i.e.\,}}
\def \eg{{\it e.g.\,}}
\def \brav #1|{\mbox{$\langle {#1}|$}}
\def \ketv #1>{\mbox{$|{#1}\rangle$}} 
\def \bracket<#1>{\mbox{$\langle {#1}\rangle$}}
\def \mate<#1|#2|#3>{\mbox{$\langle {#1}|\,{#2}\,|{#3}\rangle$}}

\section{Introduction}

Nowadays, hyperons are serious subject in physics of neutron stars.
It is naively expected that hyperons exist in core of neutron stars.
However, equation of state of matter with hyperon would be considerably soft
and seems incompatible with recent discovery of heavy neutron stars.
This is one of the most challenging puzzle in modern physics.
Because hyperon appearance in matter depends on it's chemical potential $\mu_Y(\rho$) which is given by single-particle potential $U_Y(\rho,k)$,
theoretical and experimental study of $U_Y(\rho,k)$ are very important and necessary to solve the puzzle.

Experimental studies of various hyperon-nucleus systems have been carried out and are in progress at many facilities.
For example, the first $\Xi$-hypernucleus, ${}^{14}_{~\Xi}\mbox{N}$, was found recently at KEK in Japan.
From those experimental data, single-particle potential of $\Lambda$ in nuclei is determined rather well.
While, knowledge about potential of $\Sigma$ and $\Xi$ are limited.
Theoretical studies of $U_Y(\rho,k)$ have been performed by based on some model of hyperon-nucleon interactions~\cite{Baldo:1999rq,Kohno:2009vk}.
However, there is uncertainty in the model interactions due to lack of hyperon-nucleon scattering data.

In 2006, a brand-new method was proposed to extract nucleon-nucleon interaction from QCD on lattice~\cite{Ishii:2006ec}.
This so called HAL QCD method has been successfully applied to general baryon-baryon systems~\cite{Inoue:2010hs}, meson-baryon systems and so on.
This means that we can obtain hyperon-nucleon interactions without requiring experimental data but deriving from QCD~\cite{Nemura:2008sp}.
It is very interesting to see what QCD predicts about hyperons in nuclear medium.
Therefore, in this paper, we obtain hyperon-nucleon interactions in lattice QCD simulation,
and then study single-particle potential of hyperons in nuclear matter based on the QCD interactions.

\section{Method to extract hyperon interactions from QCD on lattice}

In this section, we briefly describe how to extract hyperon interactions from QCD on lattice.
For a concrete example, let us consider interaction in strangeness $S=-1$, isospin $I=1/2$, and ${}^{1}S_0$ partial wave two-baryon sector.
In this sector, there are two flavor eigenstates $\ketv i=1,2>$, 
which are nothing but flavor irreducible representations in flavor $SU(3)$ symmetric case, 
but in general exist also in flavor $SU(3)$ broken case. 
Each eigenstate $\ketv i>$ has corresponding vector of wave function in two-baryon channel basis $\{ \Lambda N, \Sigma N \}$,
which we denote as $( \psi^{(i)}_{\Lambda N}, \psi^{(i)}_{\Sigma N} )^t$.
These wave functions should obey the Schr{\" o}dinger equation, in the Euclidean space-time, 
\small
\begin{equation}
\label{eqn:scheq}
\left( 
 \begin{array}{ccc}
 -\frac{d}{d t} &         0       \\
        0       &  -\frac{d}{d t}
 \end{array}
\right)
\left( 
 \begin{array}{c}
 \psi^{(i)}_{\Lambda N} \\
 \psi^{(i)}_{\Sigma  N}
 \end{array}
\right)
=
\left[
\left( 
 \begin{array}{cc}
     M_{\Lambda N} + \frac{-\nabla^2}{2\mu_{\Lambda N}} &  0 \\
 0 & M_{\Sigma  N} + \frac{-\nabla^2}{2\mu_{\Sigma  N}}
 \end{array}
\right)
+
\left( 
 \begin{array}{ll}
 U_{(\Lambda N)(\Lambda N)} & U_{(\Lambda N)(\Sigma N)} \\
 U_{(\Sigma  N)(\Lambda N)} & U_{(\Sigma  N)(\Sigma N)} 
 \end{array}
\right)
\right]
\left( 
 \begin{array}{c}
 \psi^{(i)}_{\Lambda N} \\
 \psi^{(i)}_{\Sigma  N}
 \end{array}
\right)
\end{equation}
\normalsize
where we've introduced a matrix of interaction potential in the coupled channel space $U_{a b}$, which we are interested in, 
and $M_{B B'}$, $\mu_{B B'}$ are the two-baryon total and reduced mass, respectively.
The potentials are non-local but energy independent in our definition.
Since the potentials are unique for the two eigenstates, they can be obtained by inverting set of these equations at each point as
\small
\begin{equation}
\label{eqn:pot1}
\left( 
 \begin{array}{ll}
 U_{(\Lambda N)(\Lambda \Lambda)} & U_{(\Lambda N)(\Sigma N)} \\
 U_{(\Sigma  N)(\Lambda \Lambda)} & U_{(\Sigma  N)(\Sigma N)} 
 \end{array}
\right)
 =
\left[
 \left( 
 \begin{array}{cc}
          \frac{\nabla^2}{2\mu_{\Lambda N}} - \frac{d}{d t} - M_{\Lambda N} & 0 \\
  0 &     \frac{\nabla^2}{2\mu_{\Sigma  N}} - \frac{d}{d t} - M_{\Sigma  N}
 \end{array}
 \right)
 \left( 
 \begin{array}{ccc}
 \psi^{(1)}_{\Lambda N} & \psi^{(2)}_{\Lambda N} \\
 \psi^{(1)}_{\Sigma  N} & \psi^{(2)}_{\Sigma  N}
 \end{array}
 \right)
\right]
\left( 
 \begin{array}{cc}
 \psi^{(1)}_{\Lambda N} & \psi^{(2)}_{\Lambda N} \\
 \psi^{(1)}_{\Sigma  N} & \psi^{(2)}_{\Sigma  N}
 \end{array}
\right)^{-1}
\end{equation}
\normalsize
where the derivative operators act inside the square bracket.
We've wrote as if $U_{ab}$ are local because we expand $U_{ab}$ in terms of local functions with the derivative operator,
and its' leading term $V_{ab}$ is just a local potential.

While, in lattice QCD simulation, we calculate the so called 4-point correlation functions
\begin{equation}
\label{eqn:4pt}
 \phi(\vec{r}, t) \equiv \frac{1}{\sqrt{Z_{B} Z_{B'}}} \sum_{\vec{x}}
                         \, \langle 0 \vert B(\vec x + \vec r,t) B'(\vec x,t) {J}(t_0)\vert 0 \rangle
\end{equation}
where $B(\vec y,t)B'(\vec x,t)$ is a product of baryon field operators at sink,
${J}(t_0)$ is a source operator which creates two baryons at $t_0$,
and $Z_{B}$ is a renormalization factor of the baryon field.
Because we can use two types of baryon operator pair, $\Lambda N$ and $\Sigma N$, at both sink and source,
we have four kinds of 4-point correlation functions $\phi_{a b}(\vec r,t)$ in this sector.
It is known that 4-point correlation function contains scattering observables, which are information of interaction,
exactly same way as quantum mechanical wave function does~\cite{Aoki:2012bb}.
The flavor eigenstate $\ketv i>$ can be exclusively generated by a source with a particular linear combination of the operator pairs.
Consequently, the wave function of the eigenstate 
is given by a linear combination of the 4-point correlation functions as
\small
\begin{equation}
\label{eqn:eigen}
\left( 
 \begin{array}{c}
 \psi^{(i)}_{\Lambda N} \\
 \psi^{(i)}_{\Sigma  N}
 \end{array}
\right)
 = 
\left( 
 \begin{array}{cc}
 \phi_{(\Lambda N)(\Lambda N)} & \phi_{(\Lambda N)(\Sigma N)} \\
 \phi_{(\Sigma  N)(\Lambda N)} & \phi_{(\Sigma  N)(\Sigma N)}
 \end{array}
\right)
\left( 
 \begin{array}{c}
 c^{(i)}_{\Lambda N} \\
 c^{(i)}_{\Sigma  N}
 \end{array}
\right) ~.
\end{equation}
\normalsize
By inserting this expression to eq.(\ref{eqn:pot1}), we get the formula which we use in our studies
to derive hadron interaction potentials from QCD on lattice in the leading order,
\small
\begin{eqnarray}
\left( 
 \begin{array}{ll}
 V_{(\Lambda N)(\Lambda N)} & V_{(\Lambda N)(\Sigma N)} \\
 V_{(\Sigma  N)(\Lambda N)} & V_{(\Sigma  N)(\Sigma N)} 
 \end{array}
\right)
&&=
\left( 
 \begin{array}{cc}
 \frac{\nabla^2}{2\mu_{\Lambda N}} \phi_{(\Lambda N)(\Lambda N)} &
 \frac{\nabla^2}{2\mu_{\Lambda N}} \phi_{(\Lambda N)(\Sigma  N)} \\
 \frac{\nabla^2}{2\mu_{\Sigma  N}} \phi_{(\Sigma  N)(\Lambda N)} &
 \frac{\nabla^2}{2\mu_{\Sigma  N}} \phi_{(\Sigma  N)(\Sigma  N)}
 \end{array}
\right)
\left( 
 \begin{array}{cc}
 \phi_{(\Lambda N)(\Lambda N)} & \phi_{(\Lambda N)(\Sigma N)} \\
 \phi_{(\Sigma  N)(\Lambda N)} & \phi_{(\Sigma  N)(\Sigma N)} 
 \end{array}
\right)^{-1}
\\
+&&
\left(
 \begin{array}{cc}
 \left(-\frac{d}{dt} - M_{\Lambda N} \right) \phi_{(\Lambda N)(\Lambda N)} &
 \left(-\frac{d}{dt} - M_{\Lambda N} \right) \phi_{(\Lambda N)(\Sigma  N)}  \\
 \left(-\frac{d}{dt} - M_{\Sigma  N} \right) \phi_{(\Sigma  N)(\Lambda N)} &
 \left(-\frac{d}{dt} - M_{\Sigma  N} \right) \phi_{(\Sigma  N)(\Sigma  N)}
 \end{array}
\right)
\left( 
 \begin{array}{cc}
 \phi_{(\Lambda N)(\Lambda N)} & \phi_{(\Lambda N)(\Sigma N)} \\
 \phi_{(\Sigma  N)(\Lambda N)} & \phi_{(\Sigma  N)(\Sigma N)} 
 \end{array}
\right)^{-1}
\nonumber
\end{eqnarray}
\normalsize
where potentials are divided into two terms, namely a Laplacian part and a time-derivative part, for convenience in analysis.
Note that $(c^{(i)}_{\Lambda N}, c^{(i)}_{\Sigma N})^t$ vanish in this expression. 
This means that we do not need the eigenstates and we can derive the potentials directly from data of the 4-point functions. 
In order to use this formula, linear independence of vector
$(\phi_{(\Lambda N)(\Lambda N)},\phi_{(\Sigma  N)(\Lambda N)})^t$ and
$(\phi_{(\Lambda N)(\Sigma  N)},\phi_{(\Sigma  N)(\Sigma  N)})^t$ is indispensable.
If the sink-source time-separation, $t-t_0$, is extremely large,
both the vectors become wave function of the ground state of sector and are linearly dependent.
However, since we always choose moderate $t-t_0$ in our calculations, this is never a problem.

Thus, we can derive potential of hyperon interactions in lattice QCD numerical simulation. 
Extension to other sector is straightforward. For example, in $S=-2$, $I=0$, ${}^{1}S_0$ two-baryon sector,
we get a $3\times 3$ matrix of potential in $\{\Lambda \Lambda, \Xi N, \Sigma \Sigma \}$ coupled channel space.

\section{Setup of lattice QCD simulation}
\label{sec:setup}

\begin{table}[t]
\caption{Mass of pseudoscalar mesons and octet baryons measured in lattice QCD simulation with the K-configuration.
 Mass of pseudoscalar mesons are taken from ref.~\cite{Ishikawa:2015rho}}
\label{tbl:mass}
\medskip
\centering
 \begin{tabular}{c|cccccc}
  \hline
  \hline
  Hadron     &  $\pi$   &   K   &   N   & $\Lambda$ & $\Sigma$ & $\Xi$ \\
  \hline
  Mass [MeV] &   146    &  525  &  956 $\pm$ 12  &   1121 $\pm$ 4   &   1201 $\pm$ 3  & 1328 $\pm$ 3  \\ 
  \hline
  \hline
 \end{tabular}
\end{table}

In order to do lattice QCD numerical simulation, we need an ensemble of gauge configuration.
Recently, the so called K-configuration set has been generated on the K computer at RIKEN AICS in Japan.
This work was done by a collaboration in HPCI Strategic Program field 5 project 1.
They employed the stout smeared Wilson clover action for quarks and the Iwasaki gauge action for gluon.
Details of the K-configuration can be found in ref.~\cite{Ishikawa:2015rho}.
One remarkable point of this ensemble is that it is generated
on a lattice with large spatial volume $V \simeq (8~\mbox{fm})^3$, 
which is very important for us to study baryon interactions.
Another remarkable point is that this full QCD gauge configuration set is generated
with 2+1 flavor dynamical quarks with almost physical mass.
Table~\ref{tbl:mass} list mass of hadrons which we obtain in our lattice QCD simulation.
One sees that measured hadron masses almost agree with the physical ones.

We adopt the so called wall type quark source for $J(t_0)$ in eq.(\ref{eqn:4pt}),
and measure the 4-point correlation functions $\phi(\vec r,t)$.
For the baryon field operator at sink, we adopt the point type one
so that we can minimize non-locality of potential $U_{a b}(\vec r, \vec r')$, 
which is our scheme to define potentials.
We put the Dirichlet boundary conditions in the temporal direction 
and take an average over forward and backward propagations in time.
We treat the non-locality by means of the derivative expansion and truncation.
We know that convergence of the expansion is fast owing to our choice of operator at sink,
and hence maintain the leading term $V_{a b}(\vec r)$ in this study.

We use 414 configurations available in the ensemble. 
In order to reduce noise and enhance signal, we repeat measurement many times for each configuration
by rotating lattice and shifting the source time $t_0$, and take the average between them.
We have done measurement $4 \times 28$ times so far, and use average of them in this study.
We still continue measuring and will have $4 \times 96$ data finally.
Therefore, the present results are not final but rather preliminary.

\section{Hyperon interaction potentials from QCD}

\begin{figure}[t]
\centering
\includegraphics[width=0.40\textwidth]{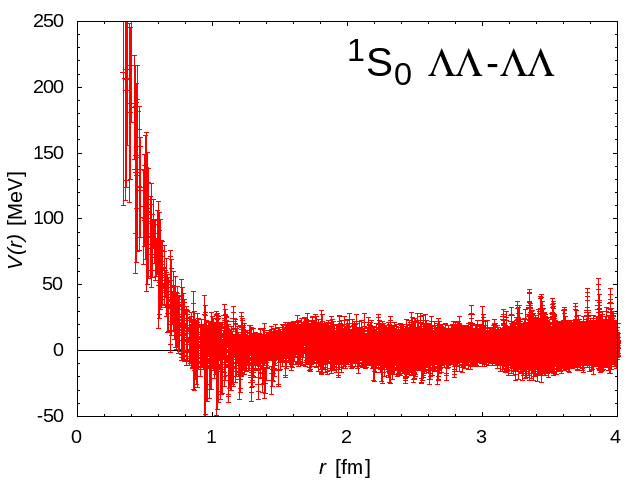} \qquad
\includegraphics[width=0.40\textwidth]{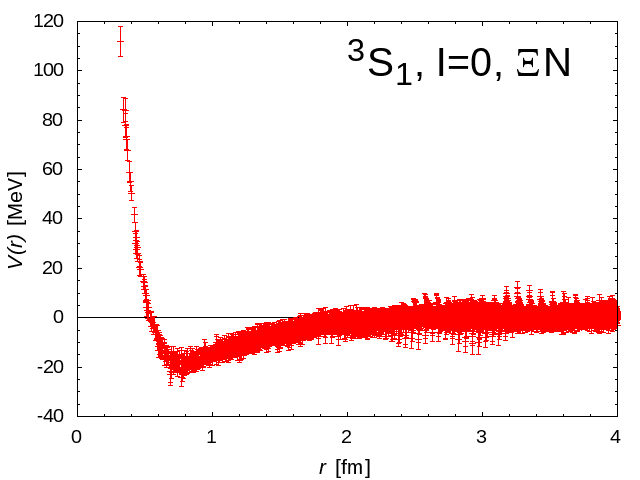}
\caption{Examples of hyperon interaction potential extracted from lattice QCD with the K-configuration.}
\label{fig:sample}
\end{figure} 

Fig.~\ref{fig:sample} shows two examples of hyperon interaction potentials $V_{ab}(r)$ obtained in our analysis.
The vertical bars show statistical error estimated by using the Jackknife method.
These potentials are obtained with the 4-point function data at sink-source separation $t-t_0=11$ in lattice unit.
In order to suppress contamination of excited baryon to the 4-point functions, we need to take a certain size for the separation.
While, signal over noise ratio of lattice QCD data become worse at large sink-source separation.
Considering the present limited number of measurement, it is most appropriate for our purpose to take $t-t_0=11$ for the moment.
We have many such hyperon interaction potentials in sectors
$S=-1,-2 \cdots$, $I=0,1/2 \cdots$, ${}^{S}L_{J} = {}^1S_0, {}^3S_1 \cdots$ all together~\cite{Doi:2015oha}. 

In general, when we apply a potential to some investigations,
we need one given in smooth function of distance or relative momentum.
Because our potential data are given at only discrete distance, we need to parameterize them before applying.
It is nervous to parameterize data with large error,
and hence it is tough to parameterize all needed potentials one by one.
Therefore, in this paper, we utilize potentials given in the flavor irreducible representation basis,
in order to reduce number of independent potentials, for the moment.
It is known that all octet-baryon pairs are classified into six independent multiplets as
$8 \times 8 = 27 + 8s + 1 + 10^* + 10 + 8a$, where the first three multiplets are symmetric
and the last three multiplets are anti-symmetric.

\begin{figure}[t]
\centering
\includegraphics[width=0.33\textwidth]{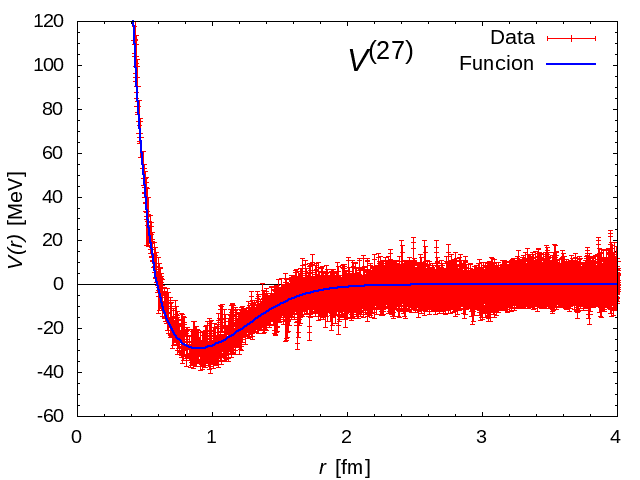}\hfill
\includegraphics[width=0.33\textwidth]{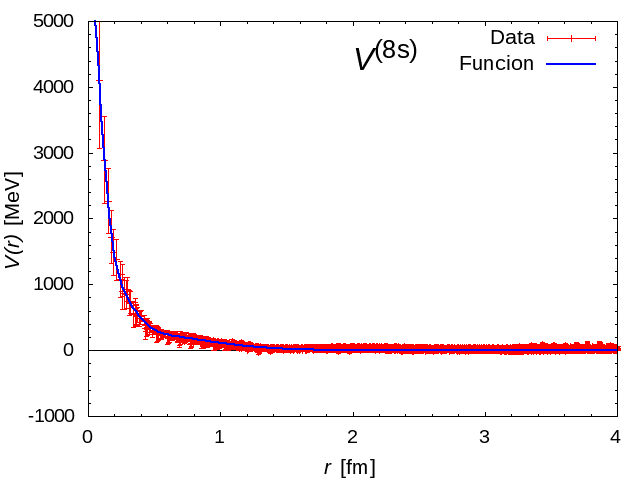}\hfill
\includegraphics[width=0.33\textwidth]{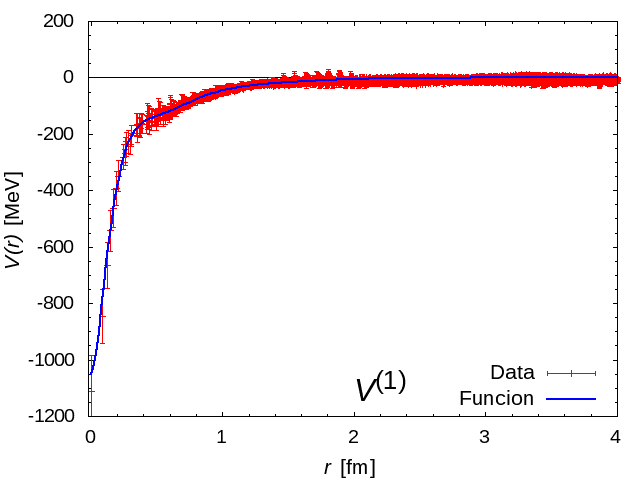}
\includegraphics[width=0.33\textwidth]{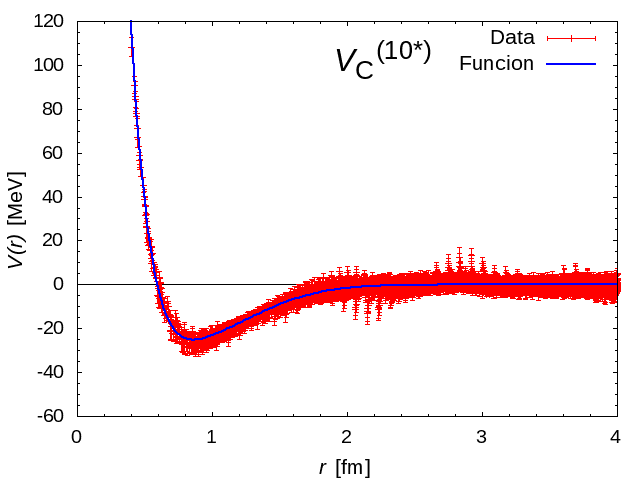}\hfill
\includegraphics[width=0.33\textwidth]{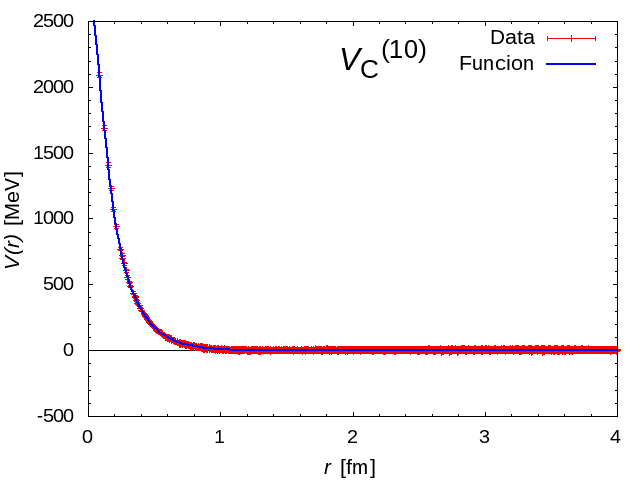}\hfill
\includegraphics[width=0.33\textwidth]{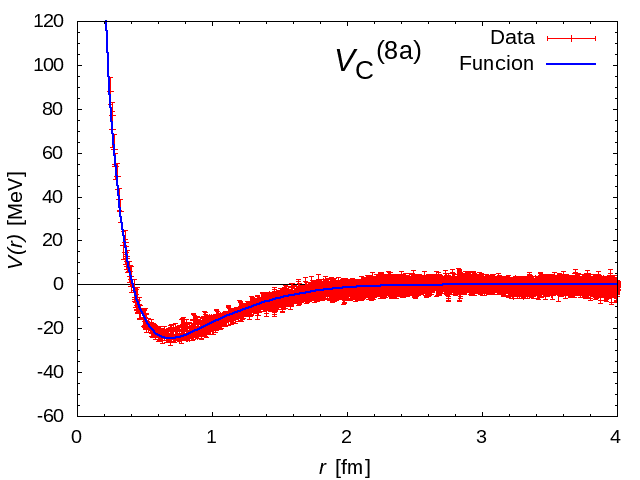}
\includegraphics[width=0.33\textwidth]{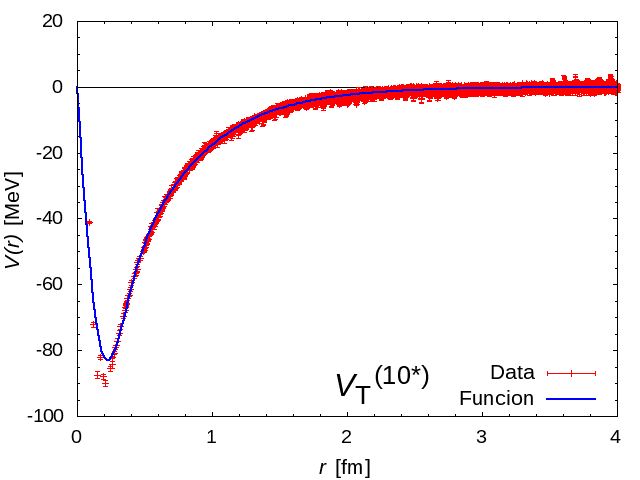}\hfill
\includegraphics[width=0.33\textwidth]{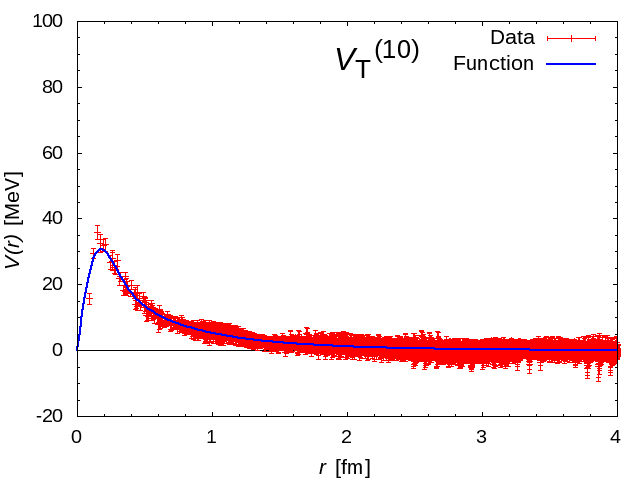}\hfill
\includegraphics[width=0.33\textwidth]{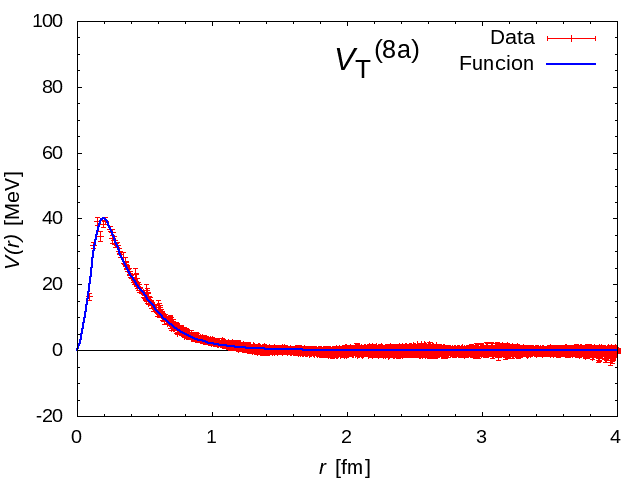}
\caption{Potential of baryon-baryon S-wave interactions in the flavor irreducible representation basis,
which are obtained by rotating data of hyperon interaction potentials in strangeness $S=-2$ sector.}
\label{fig:potentials}
\end{figure} 

We rotate data of hyperon S-wave interaction potentials in $S=-2$ sectors,
and obtain baryon S-wave interaction potentials in the irreducible representation basis.
Fig.~\ref{fig:potentials} shows obtained diagonal potentials.
Property of the diagonal potentials are same as those derived from QCD at flavor $SU(3)$ limits, for example,
entire attraction in the flavor singlet two-baryon and very strong repulsion in the flavor octet symmetric two-baryon~\cite{Inoue:2010hs}.
Blue curves show functions fitted to data in the least squares method,
where following functions are used for the central and tensor potential, respectively.
\small
\begin{eqnarray}
 V_{C}(r) &=& a_1 e^{-a_2\,r^2} + a_3 e^{-a_4\,r^2} + a_5 \left( (1 - e^{-a_6\,r^2}) \frac{e^{-a_7\,r}}{r} \right)^2
 \\
 V_{T}(r) &=& b_1 \left( 1 - e^{-b_2\,r^2} \right)^2 \left(1 + \frac{3}{b_3\,r} + \frac{3}{(b_3\,r)^2} \right)\frac{e^{-b_3\,r}}{r} ~+~
              b_4 \left( 1 - e^{-b_5\,r^2} \right)^2 \left(1 + \frac{3}{b_6\,r} + \frac{3}{(b_6\,r)^2} \right)\frac{e^{-b_6\,r}}{r}
 \nonumber
\end{eqnarray}
\normalsize

Off diagonal components are much smaller than the diagonal ones.
Since this is our first attempt to apply hyperon interactions from QCD,
we had better to start with qualitative study. 
Therefore, we omit small off diagonal components and reconstruct hyperon interaction potentials
with the diagonal components and the Clebsch-Gordan coefficients. We use them in the following.

\section{Single-particle potential of hyperons in nuclear medium}

We study hyperons in nuclear medium based on the hyperon interactions derived from QCD.
To calculate single-particle potential of hyperons in nuclear matter with density $\rho$,
we adopt the lowest order Brueckner theory, namely the Brueckner-Hartree-Fock (BHF) approximation.
In this framework, the single-particle potentials $U_{Y}(\rho,k)$ are obtained by
\begin{equation}
 \label{eqn:Uy}
 U_Y(\rho,k) = \sum_{N=n,p} \, \sum_{^SL_J} \, \sum_{k'\leq k_F}
               \mate<k,k'| G^{^SL_J}_{(YN)(YN)}(E_{YN}(k,k')) |k,k'>
\end{equation}
where $G^{^SL_J}_{(YN)(YN)}$ is G-matrix which describes $YN$ to $YN$ scattering in a partial wave ${}^{S}L_{J}$ in nuclear medium,
and solution of the Bethe-Goldstone equation with the interaction potential $V_{a b}$
\begin{equation}
 \label{eqn:BGeq}
     G_{a b}(\omega) = V_{a b} + \sum_c \sum_{k,k'}
                       V_{a c} \, \ketv k,k'> \frac{Q_c(k,k')}{\omega-E_c(k,k')+i\epsilon} \brav k,k'| \, G_{c b}(\omega)
\end{equation}
where $Q(k,k')$ is the angle averaged Pauli operator, and $E(k,k')$ is energy of baryon pair given by
\begin{equation}
 \label{eqn:Ene}
  E_{BB'}(k,k') = e_B(k) + e_{B'}(k') ~~\mbox{with}~~  e_B(k) = M_B + \frac{k^2}{2 M_B} + \mbox{Re}[U_B(\rho,k)]
\end{equation}
in the continuous choice. These highly coupled equations are solved in the iteration procedure,
and self-consistent G-matrix and $U_Y(\rho,k)$ are obtained.

\begin{table}[t]
\caption{Hyperon-nucleon coupled channels with a given total charge $Q$ and flavor symmetric $S$ or anti-symmetric $A$. 
For example, $S$ ($A$) is combined with ${}^1S_0$ (${}^3S_1$-${}^3D_1$) partial wave, in this study} 
\label{tbl:channel}
\medskip
\centering
\small
 \begin{tabular}{l|c|c|c|c}
  \hline
  \hline  
                 &  $Q=0$    &   $Q=+1$  &   $Q=-1$   & $Q=+2$ \\
  \hline
  $(YN)_{S,A}$   & $\Lambda  n$, $\Sigma^0 n$, $\Sigma^- p$ & $\Lambda  p$, $\Sigma^0 p$, $\Sigma^+ n$ & $\Sigma^- n$ & $\Sigma^+ p$ \\
  \hline
  $(\Xi N)_{S}$  & $\Xi^0 n$, $\Xi^- p$,  $\Sigma^+ \Sigma^-$, $\Sigma^0 \Sigma^0$, $\Sigma^0 \Lambda$, $\Lambda \Lambda$ &
                   $\Xi^0 p$, $\Sigma^+ \Lambda$  & $\Xi^- n$, $\Sigma^- \Lambda$ & \\
  \hline 
  $(\Xi N)_{A}$  & $\Xi^0 n$, $\Xi^- p$, $\Sigma^+ \Sigma^-$, $\Sigma^0 \Lambda$ &  $\Xi^0 p$, $\Sigma^+ \Sigma^0$, $\Sigma^+ \Lambda$ &
                   $\Xi^- n$, $\Sigma^- \Sigma^0$, $\Sigma^- \Lambda$  & \\
  \hline
  \hline
 \end{tabular}
\normalsize
\end{table}

Recall that hyperon-nucleon interactions connect many channels.
Table~\ref{tbl:channel} list the coupled channels which enter in this calculation.
We truncate the partial wave expansion and include only ${}^1S_0$ and ${}^3S_1$-${}^3D_1$,
because we have not derived hyperon interactions in odd parity partial waves.
This truncation must be reasonable at low density.
For baryon mass $M_B$, we use the physical values, because we want to focus on the hyperon interactions,
and baryon masses in our QCD simulation are almost physical.
For the single particle potentials of nucleon $U_p(\rho,k)$ and $U_n(\rho,k)$, 
we use ones obtained with the AV18 phenomenological two-nucleon force in the BHF approximation.

\begin{figure}[t]
\centering
\includegraphics[width=0.4\textwidth]{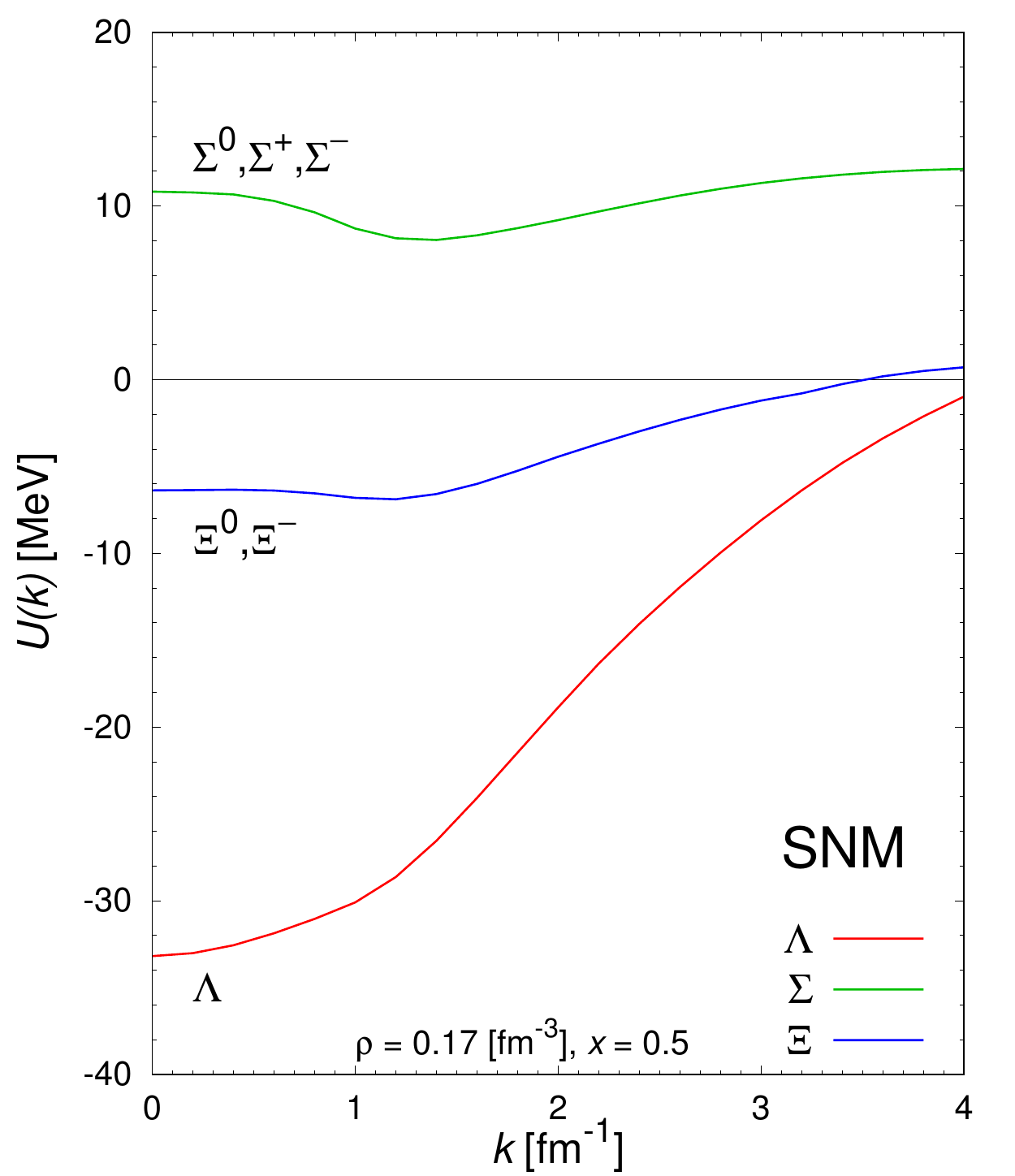} \qquad
\includegraphics[width=0.4\textwidth]{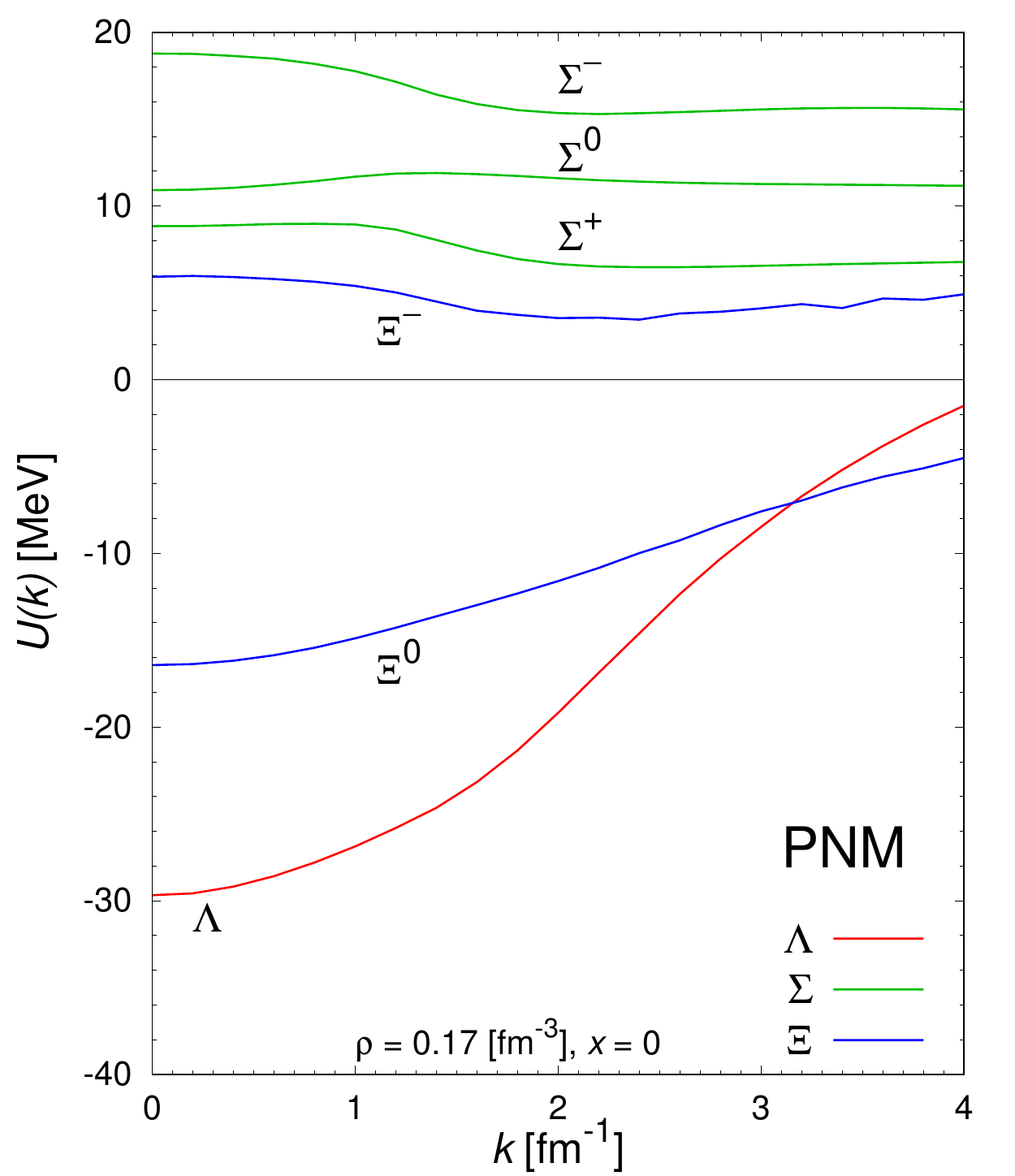}
\caption{Hyperon single-particle potentials in nucleonic matter at the normal nuclear density,
which are obtained by based on the hyperon interactions from QCD on lattice.
Note that these are preliminary because we have not finished whole lattice QCD measurement
and not estimated influence of statistical error yet.}
\label{fig:uys}
\end{figure}

We calculate $U_Y(\rho,k)$ in two representative nuclear matter, namely,
in the symmetric nuclear matter (SNM) and the pure neutron matter (PNM).
Fig.~\ref{fig:uys} shows real part of obtained $U_Y(\rho_0,k)$ in SNM and PNM,
where $\rho_0=0.17$~fm$^{-3}$ is the normal nuclear matter density.

Because matter at center of heavy nuclei is analogous to SNM, 
information of $U_Y(\rho_0,0)$ in SNM can be obtained through hypernuclear experiment.
Current experimental data seems to indicate that 
$U_{\Lambda}(\rho_0,0) \simeq -30$~MeV, $U_{\Sigma}(\rho_0,0) \geq 20$~MeV, and  $U_{\Xi}(\rho_0,0) \simeq -10$~MeV.
Because of much data of $\Lambda$-hypernuclei, $U_{\Lambda}(\rho_0,0) \simeq -30$~MeV is rather established.
While, experimental data on $\Sigma$ and $\Xi$ are poor and the above indications are need to be refined and confirmed.
We see that our QCD based hyperon interactions predict that
$U_{\Lambda}(\rho_0,0)=-33$~MeV, $U_{\Sigma}(\rho_0,0)=11$~MeV, and $U_{\Xi}(\rho_0,0)=-6$~MeV for SNM,
which are almost consistent with the experimental data. This agreement, especially for $\Lambda$, is remarkable.
Recall that we have not used any phenomenological input for hyperon interactions but used only QCD.
Hence, we can recognized this is the first-ever explanation of a hypernuclear physics based on QCD.
Moreover, one can regard that the less certain experimental information of $U_{\Sigma}(\rho_0,0)$ and $U_{\Xi}(\rho_0,0)$ are qualitatively supported by QCD.

Because large part of neutron star core consist of high density PNM,
theoretical prediction of $U_Y(\rho,k)$ in PNM is useful to study hyperons in neutron star and interesting. 
In Fig.~\ref{fig:uys}, we see that general feature of predicted $U_Y(\rho_0,k)$ in PNM is same as that in SNM,
namely, $\Lambda$ get attraction and $\Sigma$ get repulsion.
One remarkable point is that our lattice QCD hyperon interactions predict
that $U_{\Xi}(\rho_0,k)$ in PNM depends on $\Xi$'s change strongly so that $\Xi^0$ get attraction but $\Xi^-$ get repulsion in PNM.
This result suggests that more neutron rich nuclei are more advantageous to bind $\Xi^0$,
while symmetric nuclei would be most advantageous to bind $\Xi^-$.
It is interesting if we could verify this by experiment in future.

\section{Summary and outlook}

In this paper, we've calculated hyperon single-particle potentials $U_Y(\rho,k)$ in nuclear medium starting from QCD on lattice.
First we've extracted hyperon interaction potentials from QCD by means of the HAL QCD method, and then applied them to the BHF approximation.
Remarkably, obtained $U_Y(\rho_0,k)$ for SNM are consistent with experimental data.

In this study, we have not discussed uncertainty of $U_Y(\rho,k)$ which comes from statistical error of the lattice QCD hyperon interaction potentials.
We guess it is order of 10\% from our experience, but we will estimate it properly in our next paper.
We've used rotated potential data and omitted off-diagonal components in order to reduce number of painful parameterization.
We will use original potential data when we finish whole measurement so that we can include physical explicit flavor $SU(3)$ breaking.
We've truncated the partial wave expansion in eq.(\ref{eqn:Uy}) at $^3\mbox{S}_1$-$^3\mbox{D}_1$ due to lack of interaction potentials.
This truncation must be good at $\rho_0$ but will not be valid at higher density.
Because we are interested in chemical potential of hyperons in high density matter, we need to predict $U_Y(\rho,k)$ at high density.
Therefore, it is our next target to extract hyperon interactions in higher partial waves from QCD on lattice.

\section*{Acknowledgments}
The author thanks the PACS Collaboration for generating and providing gauge configurations,
and the JLDG team~\cite{JLDG} for providing storage to save our data. 
Numerical computation of this work was carried out on 
the K computer at RIKEN AICS (hp120281, hp130023, hp140209, hp150223, hp150262, hp160211), 
the HOKUSAI FX100 computer at RIKEN Wako (G15023, G16030),
and the HA-PACS at University of Tsukuba (14a-20, 15a-30).
This research is supported in part by the JSPS Grant-in-Aid for Scientific Research (C)26400281,
Strategic Program for Innovative Research (SPIRE) Field 5 project, 
and "Priority Issue on Post-K computer" (Elucidation of the Fundamental Laws and Evolution of the Universe).

\end{document}